# Gas Dynamic Virtual Nozzle for Generation of Microscopic Droplet Streams

**D.P. DePonte, U. Weierstall, D. Starodub, K. Schmidt, J.C.H. Spence, and R.B. Doak**

Department of Physics, Arizona State University, Tempe, AZ  85287-1504, USA

**Abstract**
As shown by Gañán-Calvo and co-workers, a free liquid jet can be compressed in diameter through gas-dynamic forces exerted by a co-flowing gas, obviating the need for a solid nozzle to form a microscopic liquid jet and thereby alleviating the clogging problems that plague conventional droplet sources of small diameter.  We describe in this paper a novel form of droplet beam source based on this principle.  The source is miniature, robust, dependable, easily fabricated, and eminently suitable for delivery of microscopic liquid droplets, including hydrated biological samples, into vacuum for analysis using vacuum instrumentation.  Monodisperse, single file droplet streams are generated by triggering the device with a piezoelectric actuator.  The device is essentially immune to clogging.

**Introduction.**

Fabrication and analysis of solid materials often takes place under high vacuum (HV) or ultra-high vacuum (UHV) conditions.  A vacuum environment is vital if the sample is reactive or if the deposition/probe beam (electrons, ions, molecules, soft x-rays, IR radiation, etc.) can be scattered, absorbed, or chemically degraded by an ambient atmosphere.  Consequently, a huge family of UHV fabrication and diagnostic techniques has emerged over the past forty years and now finds widespread use in materials science and technology.  Use of these techniques has historically been restricted to non-volatile samples.  Since evaporative flux from a liquid surface scales with surface area, however, volatility need not be an absolute limitation.  Liquid samples, including biological species immersed in aqueous solution, can be compatible with HV or even UHV conditions provided the sample is a liquid droplet of sufficiently small size, generally microscopic.  Small droplet size is also imperative if the probe beam has only limited penetration [1], e.g., soft x-rays (~1 μm at 1 keV) or electrons (~100 nm at 100 keV), or if the probe beam itself has only microscopic dimensions.  The latter is notably the case for the emerging generation of free-electron lasers (beam cross sections of 70 μm x 70 μm for the European XFEL [2] and 10 μm x 30 μm for the LCLS [3]).

With this in mind, a monodisperse single-file train of neutral microscopic droplets becomes a highly attractive means of injecting hydrated biological samples into vacuum for study using UHV instruments.  Rayleigh break-up of a liquid jet delivers exactly this form of droplet stream [4], but invariably limited by nozzle clogging to droplet diameters of ~20 μm or greater, which is still too large for penetration by electrons and soft x-rays.   Much smaller droplets (~1 μm diameter) can be produced by electrospray or electrospray-assisted Rayleigh sources [1], but only in an electrically charged form.  Since droplet charging may denature a biological molecule contained within the droplet, electrospray sources are a less preferred choice if structure determination of the molecule is the experimental objective [5].

Our particular interest is the use of microscopic aqueous droplets to deliver proteins into vacuum for serial diffraction of by x-rays or electrons [6, 7].  Such studies are facilitated by aligning the proteins via polarization in an intense optical field [8].   Both protein alignment and probe transmission improve dramatically with decreasing droplet size and it appears that a hydration shell of only a few monolayers thickness suffices to maintain the native conformation of a protein [9]. Accordingly, a major goal of our droplet generation efforts is to produce the smallest electrically neutral droplets possible, preferably much less than 1 μm diameter. For many applications, it is advantageous that the droplets be delivered as a periodic straight-line stream of monodisperse droplets, locked to the phase and frequency of an external trigger signal. In this manuscript we describe the development and testing of a novel droplet beam source that does exactly this, while essentially eliminating clogging as an experimental concern.

**Background**

Driven by a reduction in surface free energy, any free cylindrical jet of liquid emerging in laminar flow from an orifice breaks up spontaneously up to form a train of spherical droplets. This instability was first analyzed in the 1800's, notably by Rayleigh [10, 11] whose name it now bears. Seminal work on microscopic liquid jets in vacuum was carried out by Faubel and co-workers, who investigated jets of pure water [12 - 13], organic solvents [14, 14], and various mixtures of liquids [15]. They reported [12] stable laminar flow of water from 5 to 20 μm diameter nozzles at Reynolds numbers [16] ranging from $Re$ = 250 to 1200. Spontaneous Rayleigh break-up yields a narrow but not monodisperse size distribution of droplets. By applying a small periodic excitation near the spontaneous break-up frequency (e.g. acoustic oscillations from a piezoelectric device), Rayleigh break-up can be "triggered" to lock droplet production to the frequency and phase of the excitation signal [1,4]. The droplet stream then becomes perfectly periodic and monodisperse.

All critical flow dimensions of spontaneous Rayleigh break-up scale with the diameter $D$ of the nozzle [17], including the mean break-up segmentation length (*4.55 D*), the mean droplet diameter (*1.90 D*), and the distance from the nozzle to onset of break-up (13 $We^{1/2}D$, where $We$ is the Weber number [18]). Given the scaling of droplet diameter with nozzle diameter it is therefore possible, in principle, to generate droplets of arbitrarily small size simply by reducing the nozzle diameter. However, nozzle clogging becomes a major complication for $D$ < 10 μm [19], usually limiting attainable droplet size to ~20 μm or larger.

In the late 1990's, Gañán-Calvo and collaborators [20] introduced a very different method of generating very small diameter columnar liquid jets. They positioned a straight-walled capillary tube (0.80 mm diameter, typically) just upstream of an aperture (150 μm diameter, typically) in a flat plate that formed the end of a secondary plenum surrounding the capillary. A gas was passed through this secondary plenum, flowing coaxially about a liquid jet emerging from the capillary in such a fashion that the liquid jet and coaxial gas flow passed together in laminar flow through the aperture, driven by a pressure differential across the aperture. The forces exerted on the liquid stream by the coaxial gas stream caused the liquid jet to neck down into a linear "microthread" of much smaller diameter than either the capillary or the aperture. Gañán-Calvo reported microthread diameters ranging from roughly 10 to 200 μm depending on the liquid density and viscosity, gas pressure difference across the aperture, and liquid flow rate [21, 21].

The use of fluid dynamic forces to compress the diameter of a liquid jet is well known in the field of microfluidics [22], but apparently always limited prior to the Gañán-Calvo work to the use of an immiscible coaxial liquid sheath as the compressing medium. The demonstration that a coaxial gas sheath could effect the same compression was therefore a major advance. In essence the coaxial gas-dynamic flow of the Gañán-Calvo design constitutes a "virtual" nozzle, forcing a reduction in the diameter of the entrained liquid jet just as would a solid-walled nozzle. Accordingly, we refer to this as a Gas Dynamic Virtual Nozzle (GDVN). The coaxial flow, being gaseous, can easily be pumped away via differential pumping downstream of the GDVN aperture, yielding a pure liquid stream surrounded at most by its own vapor.

A key aspect of the Gañán-Calvo method is that the forced reduction in diameter of the liquid jet occurs over a very short distance, specifically over a distance shorter than the gestation length for Rayleigh break-up. Consequently it is the narrow microthread that undergoes Rayleigh break-up rather than the larger diameter parent jet, yielding droplets that are much smaller than would be achieved by Rayleigh break-up of the parent jet itself. Equally significant is that, in contrast to a solid-walled nozzle, the inner capillary tube of the Gañán-Calvo can be completely straight-walled with no convergent section to trap contaminant particles. The smallest mechanical opening is that of the capillary or of the flat-plate aperture, both of which can be much larger in diameter than the liquid microthread produced by the gas dynamic compression. A contaminant particle smaller than either of these two openings simply passes through the source system with nothing more than a momentary disruption of the flow field. The Gañán-Calvo design is essentially immune to clogging.

**Adapting the Gañán-Calvo Design**

With these attributes, some form of GDVN was an obvious candidate droplet source for our serial diffraction measurements in vacuum. However, our desired droplet size of 0.1 to1 μm diameter was much smaller than had ever been produced by the Gañán-Calvo group and hence it was necessary to demonstrate that the Gañán-Calvo scheme still functioned in this smaller size regime. Moreover, our intended usage in HV or UHV required that the GDVN design be adapted for vacuum compatibility. For both of these reasons, we constructed our version of GDVN with much smaller capillary and gas apertures (20-50 μm range) than in the Gañán-Calvo design and operated it at lower flow rates. Mechanical alignment of the capillary exit with the GDVN aperture becomes a significant experimental issue with openings this small, compounded in our case by the need to transmit any *in situ* alignment motion through a vacuum wall.

Our first variety of GDVN, shown in Fig. 1, was essentially a scaled down version of the original Gañán-Calvo design. The exit of a thick-walled capillary tube was positioned facing an orifice in the flat end-wall of a large plenum through which the coaxial gas flowed. The glass wall of the plenum allowed the aerodynamics of the flow to easily be observed and photographed. The outer wall of the capillary was tapered, as is evident in the figure, to facilitate laminar flow of the gas around the front edge of the capillary and to minimize the base diameter of the emergent liquid jet, which would wet the entire front end of the capillary tube under certain operating conditions. The inner bore of the capillary was of constant diameter, as may be seen by careful scrutiny of Fig. 1. A piezoelectric tilter combined with an inchworm drive provided *in situ* transverse and axial positioning, respectively, of the capillary exit relative to the aperture. This allowed very high alignment accuracy as seen in Fig. 1, which shows a 20 μm ID capillary tube aligned with a 30 μm diameter aperture. The liquid jet is visible emerging from the bore of the capillary and forming a conical cusp as it is compressed in diameter upon passage through the GDVN aperture. Experiments with this source indicated that the flow dynamics of the Gañán-Calvo would indeed scale down to at least micron droplet sizes.

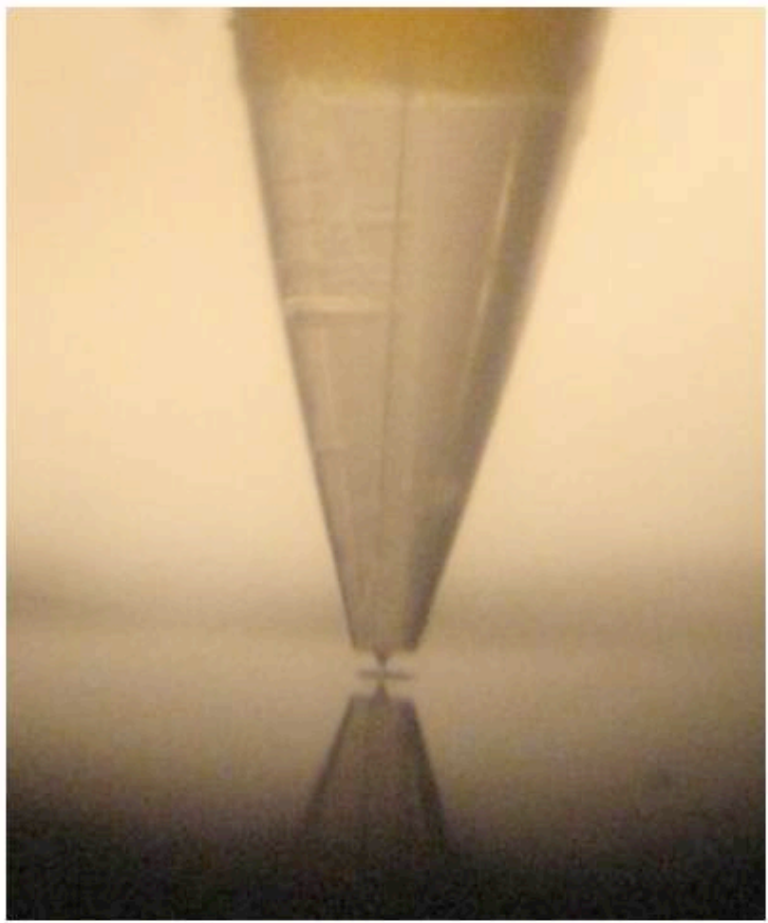

**Fig. 1** – Scaled-down version of the Gañán-Calvo nozzle as constructed at ASU. A liquid jet emerges from a straight-bore capillary tube (360 μm OD, 20 μm ID nozzle, tapered outer wall) and passes with a coaxially flowing gas flow through a 30 μm ID aperture in a flat plate, compressing the jet in diameter to form the conical liquid cusp that is seen both directly and as a reflection in the plate. Piezoelectric drives on the capillary tube allow its exit to be accurately positioned with respect to the aperture.

Our second variety of GDVN represented a considerable departure from the Gañán-Calvo scheme. In this source, shown in Fig. 2, the aperture in a flat plate was replaced with an aerodynamically converging sidewall in a miniature (~1 mm diameter) glass tube through which the coaxial gas flowed. The inner capillary tube was centered within this outer housing by means of a tubular sleeve of appropriate ID and OD, the alignment therefore being entirely passive. Since the components were transparent, the liquid flow could easily be observed and photographed at all points along the flow axis. The small size of this source greatly facilitates incorporation of differential vacuum pumping to reduce the gas load on the vacuum chamber. This second, miniature GDVN design proved to be simple, robust, and reliable. It will be the main subject of this paper.

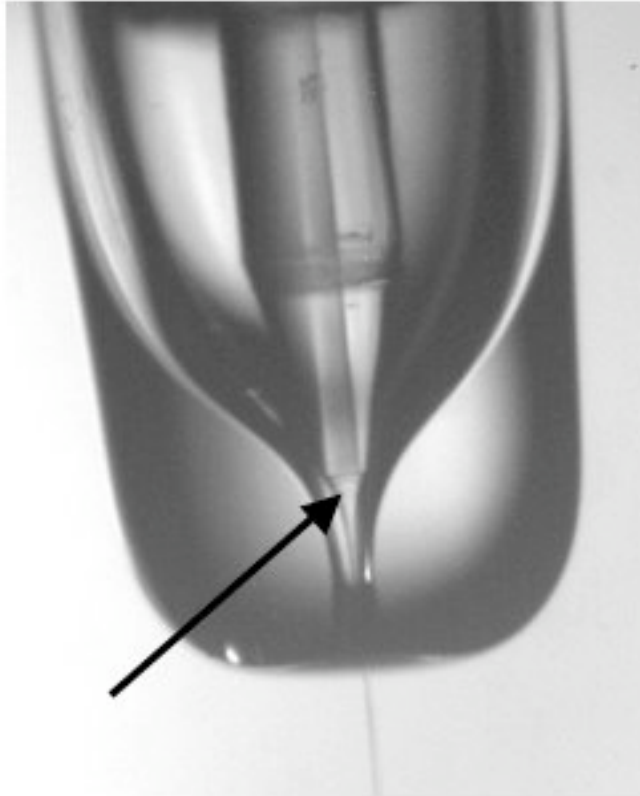

**Fig. 2.** – View of the exit end of the miniature version of gas dynamic virtual nozzle, photographed in operation with a water jet (arrow) emerging from the central capillary (360 μm OD, 50 μm ID, tapered outer wall) to be compressed by gas dynamic forces as the liquid stream passes with a co-flowing coaxial gas flow through the exit channel of the outer plenum (1.2 mm OD). A PTFE sleeve that centers the capillary within the outer housing is just out of view at the top of the photograph.

**Fabrication of the Miniature ASU GDVN**

The miniature GDVN of Fig. 2 incorporates a commercial hollow-core fused silica optical fiber (360 μm OD and 20-50 μm ID; Polymicro Technologies LLC) as the inner capillary (liquid channel). A commercial borosilicate glass capillary (1.2 mm OD by 0.9 mm ID; Sutter Instrument) forms the outer housing (gas plenum). To form a exit channel on the gas plenum, the borosilicate tube is held vertically and rotated about its axis as the tube end is heated from below with a standard propane torch. The sidewall thickens at the heated end to form the radially symmetric, convergent exit channel seen in Fig. 2 [23]. This replaces the thin, sharp-edged, flate-plate aperture of the original Gañán-Calvo design. Apart from its smoothly-varying aerodynamic sidewall, the gas dynamic "aperture" is now an actual channel, with an aspect ratio (length to diameter) of much greater than unity. The exact shape of the sidewall and channel can be varied by pressurizing the tube during heating (as in pressure polishing [24]) and by post-formation grinding back of the front end (as is often done in fabricating standard Rayleigh droplet sources [24]). The performance of the GDVN, at least for liquid jets emerging into stagnant air, seems to be rather insensitive to these shapes.

An external taper was again cut onto the exit end of the liquid capillary, as is easily observed in Fig. 2. This was done using a 12 µm grit grinding disk on an Allied TechPrep polishing machine. While grinding, the silica fiber was held in contact with the grinding disk by means of a custom jig that allowed the taper angle to be set as desired and, in addition, the fiber to be rotated about its axis to promote formation of a symmetric taper. Alternatively, commercial capillary tubes can be ordered with an external taper already ground onto the end [25].

A 2 cm long sleeve cut from commercial PTFE tube (Small Parts Inc.) centered the liquid capillary tube within the outer gas plenum. Transverse alignment of the capillary end with the GDVN exit channel was thereby entirely passive, established by the fit of the capillary tube inside the sleeve and the sleeve inside the outer housing. A tighter fit produced better alignment, but this was limited both by the need to slide the capillary through the sleeve to adjust the axial position of the liquid jet and by the need for the coaxial gas flow to make its way past the sleeve. When necessary, additional clearance for the gas flow was achieved by carefully shaving down the sleeve in thickness at two or more locations along its periphery. This was generally necessary only for small gas apertures ( <50 µm ID) where alignment was critical, making a tight press fit of the sleeve in the outer tube mandatory. For gas apertures with a diameter larger than 50 µm, a fairly loose sliding fit of the sleeve in the outer tube provided adequate clearance for the gas flow without compromising the transverse alignment. Actual dimensions of the sleeve material, as measured in the lab, often exceeded the manufacturer's published tolerances and made tedious matching of individual components compulsory prior to assembly.

With the alignment sleeve in place, the inner and outer tubes were positioned axially to give a desired separation between capillary exit and GDVN channel. A 100 µm ID capillary tube was inserted into the distal end of the gas plenum, providing a connection through which gas could be supplied to the plenum, and the tubes were then permanently glued together and sealed with a drop of epoxy at this junction. This gave a robust self-contained unit as shown in Fig. 3. Alternatively, the outer housing could be mounted in the straight-run of a standard HPLC 3-way cross. The outer capillary was then terminated in the cross (allowing gas to be supplied to the gas plenum through the side run) and the inner capillary passed through the cross to be clamped on the far end of the straight run. The capillary-exit separation could then be adjusted by loosening the latter connection and sliding the inner capillary axially with respect to the outer housing.

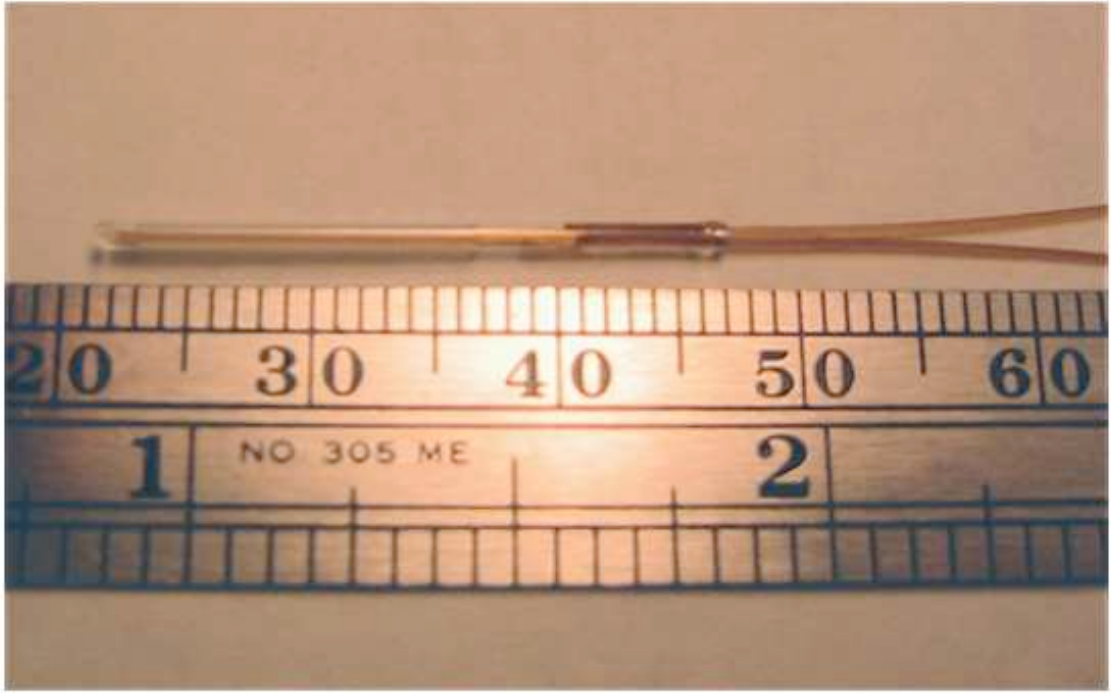

**Fig. 3** - Miniature GDVN system.  The exit is at the left.  The ruler scales are mm (top) and inches (bottom).  The centering sleeve can just be discerned in this image, extending from about 23 to 37 mm.  A drop of epoxy at the distal end (right) seals the 1.2 mm OD outer glass housing about the liquid capillary (lower tube, 360 μm OD) and the gas supply (upper tube, also 360 μm OD).

Free-running droplet beam sources can often be triggered by applying a periodic acoustic signal at a frequency near the spontaneous break-up frequency.  To this end, a small piezoelectric actuator could be clipped to the outside of the outer glass tube of the miniature GDVN.  The piezoelectric actuator and its drive electronics were exactly the same as used successfully in our previous studies of conventional Rayleigh droplet beams [1].  It was by not obvious at the outset, however, that droplet generation could be triggered in the GDVN in this fashion:  The applied acoustic signal could only reach the liquid jet circuitously, either traveling through the gas flow surrounding the liquid jet or via a long mechanical pathway to the rear of the outer tube and then back forwards through the inner tube.

**Operation**

The miniature GDVN is shown in operation in Fig. 2.  The PTFE sleeve that centers the inner capillary tube within the outer glass housing lies just above the top of the photograph and so is not seen in this photograph.  Sample liquid was supplied to the inner capillary via either a syringe pump (low pressure operation) or a gas-pressurized liquid reservoir (high pressure).  The liquid jet emanating from the 50   μm ID inner tube is accelerated by the gas flowing through the surrounding outer tube and necks down to exit the GDVN channel with a much smaller diameter than that of the liquid supply tube.  Accordingly, gas dynamic compression is seen to work quite effectively even in this very different geometry from the original Gañán-Calvo design.

Microfluidic devices generally exhibit rather complex flow behavior as a function of drive pressure [26] and our miniature GDVN was no exception, with both gas pressure and liquid pressure playing a role.  Three principal regimes of behavior were observed:

(1) "Dripping" - At low liquid and low gas pressures, large single drops were emitted from the GDVN channel, varying from 5 to 50 μm in diameter.  Doublets or higher order multiplets of droplets were emitted under certain operating conditions, often in such a fashion that the individual multiplet droplets coalesced further downstream.  The details of these emission and coalescence events could be extremely regular from one to the next.

(2) "Spurting" - At higher gas pressure but still low liquid pressure, a long, slender column of liquid was periodically emitted.  This column then broke up in flight via Rayleigh instability to yield a finite linear train of droplets.

(3) "Jetting" - At still higher liquid pressure, a continuous microthread of liquid emerged and underwent Rayleigh break-up to yield a continuous, single-file train of miniature droplets.  This was the desired mode of operation.  The pressures to reach this regime were beyond the capacity of a syringe pump and so required use of the gas-pressured liquid reservoir.  With a 50 μm ID capillary of 50 cm length from reservoir to capillary exit, 250 psi at the liquid reservoir was typically required to reach this regime.  Even higher pressures were needed for jetting from smaller diameter or longer tubes.

There was considerable hysteresis in the squirting-to-jetting transition, the transition taking place at higher values as the liquid pressure was being raised than when it was being lowered.  In liquid-liquid microfluidic experiments [27], a dripping-to-jetting transition has been shown to take place at $We$~3 over a wide range of capillary numbers.

Operation at too low or too high a gas pressure yielded unsatisfactory behavior regardless of liquid pressure.  Gas dynamic compression clearly must fail at overly low gas pressure, which allows the liquid emerging from the inner capillary to fill the entire GDVN exit channel.  At very high pressure the liquid jet would come into contact with the sidewall of the exit channel – presumably due to Venturi or inertia effects – and this also disrupted the flow.

**Experimental Results.**

We have described elsewhere an optical microscopy system for recording fast single-shot images of droplet streams [1].  This same system was employed to image the droplet trains generated by the GDVN under various operating conditions.  Several such images are shown in Figs. 4 and 5.  Of particular interest was the observation that the GDVN could indeed be "triggered" by an acoustic vibration applied to the outer glass tube.  This is illustrated in Fig. 4 for operation in the jetting regime with (a) spontaneous break-up in the absence of an applied acoustic vibration, (b) break-up in the presence of a 73 KHz vibration, and (c) break-up in the presence of a 169 kHz vibration.  Untriggered, the initial columnar jet extends beyond the exit of the GDVN channel as observed in Fig. 4(a).  With the acoustic trigger signal applied, Figs. 4(b) and (c), the break-up point moves upstream into the GDVN channel and the droplet train becomes monodisperse and periodic.  The spacing and size of the droplets varies accordingly, as dictated by continuity for a given flow velocity.  Triggering in this manner was possible only in the jetting regime and only for low gas pressures.  In the dripping regime, triggering was not possible nor was it possible to produce a uniform droplet size.

Also of considerable interest was the variation in flow morphology as the driving pressure of the coaxial gas flow was increased.  This is illustrated in Fig. 5.  At low gas pressures, Fig. 5(a), the droplet diameter was roughly twice that of the columnar parent jet, consistent with Rayleigh

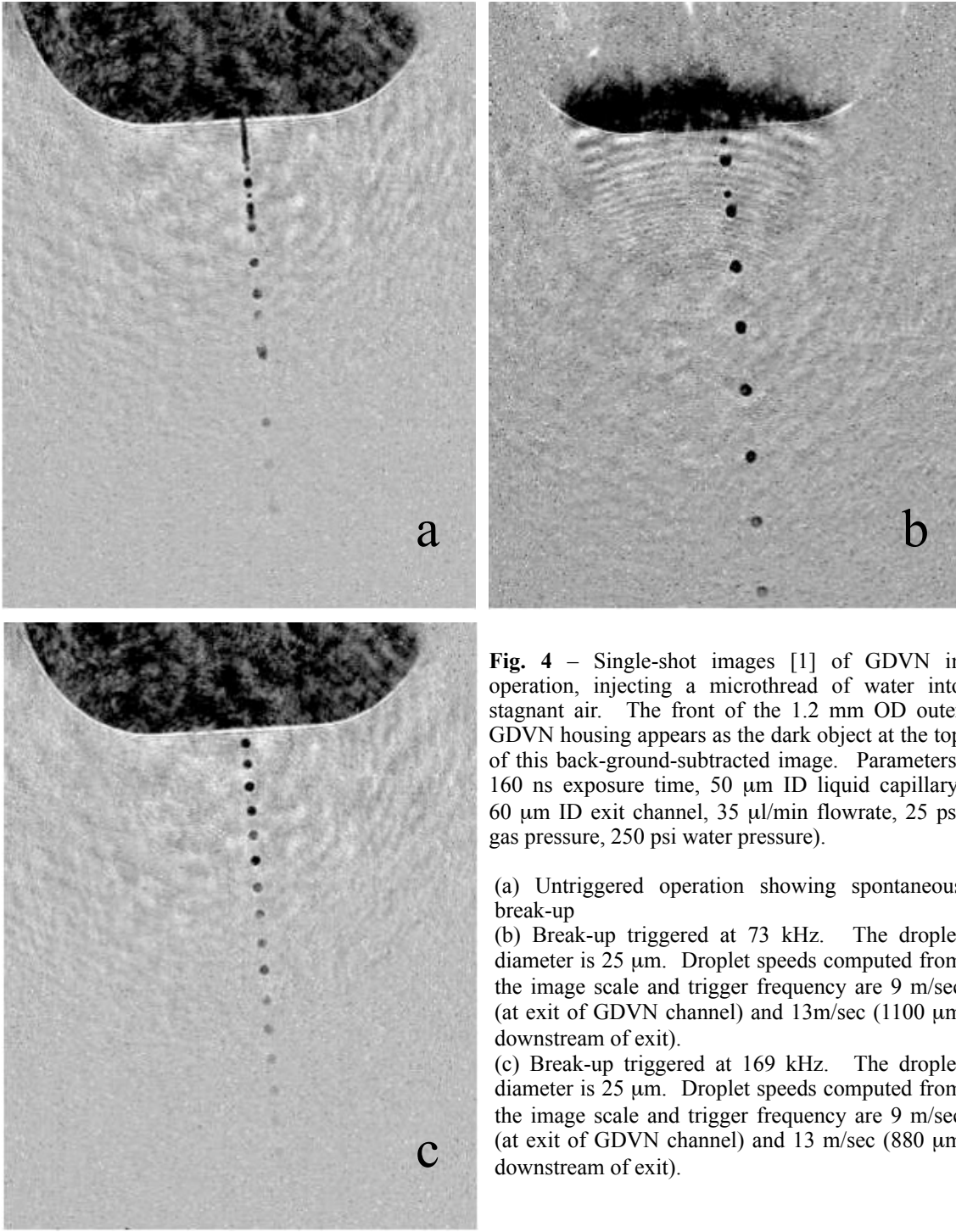

Fig. 4 – Single-shot images [1] of GDVN in operation, injecting a microthread of water into stagnant air. The front of the 1.2 mm OD outer GDVN housing appears as the dark object at the top of this back-ground-subtracted image. Parameters: 160 ns exposure time, 50 μm ID liquid capillary, 60 μm ID exit channel, 35 μl/min flowrate, 25 psi gas pressure, 250 psi water pressure).

(a) Untriggered operation showing spontaneous break-up
(b) Break-up triggered at 73 kHz. The droplet diameter is 25 μm. Droplet speeds computed from the image scale and trigger frequency are 9 m/sec (at exit of GDVN channel) and 13 m/sec (1100 μm downstream of exit).
(c) Break-up triggered at 169 kHz. The droplet diameter is 25 μm. Droplet speeds computed from the image scale and trigger frequency are 9 m/sec (at exit of GDVN channel) and 13 m/sec (880 μm downstream of exit).

break-up triggered at about the spontaneous break-up frequency. At high gas pressure (high *We*) this was no longer the case; rather the droplet diameter was seen, Fig. 5(b), to be roughly equal to the jet diameter. As discussed by Gañán-Calvo and co-workers [27], this is likely the result of shear forces arising at the free boundary of the liquid jet when operating at the higher gas velocities. These forces become the dominant driving force, and triggering by application of an external acoustic signal is no longer possible.

9 of 12

The miniature GDVN of Fig. 2 has been successfully operated under HV conditions by surrounding the nozzle with a differential pumping plenum. The small size of the device greatly facilitates this. In fact, the entire GDVN system of Fig. 2 can replace the single capillary of Fig. 1, with the gas plenum of that source being used as differential pumping stage and the droplet beam from the GDVN exiting through the flat-plate orifice into vacuum. Alternatively, a condensable gas may be used as the GDVN coaxial gas flow. Surrounding the nozzle with liquid nitrogen-cooled panels then provides cryo-pumping of very high pumping speed. We have used both approaches successfully, yielding vacuua of $10^5$ torr and below in a 10 l vacuum chamber pumped at 500 l/s. When run in vacuum, the exit of the GDVN is cooled by the free expansion of the outflowing coaxial gas. This can lead to ice formation in the GDVN channel if the liquid jet momentarily contacts exit channel wall, for example on startup when air bubbles in the liquid line disrupt the liquid flow. Heating the nozzle to remove the ice generally restores normal operation.

We have not yet determined exactly how small a droplet can be produced with our GDVN. The device appears to run in a mode in which liquid is passing out of the nozzle and can be collected downstream, yet no droplets are seen in an optical microscope. This would be the case if droplets were too small to be resolved by visible light. We have very recently run our GDVN successfully in a scanning electron microscope (SEM), imaging the droplets via electron scattering rather than visible light, and hope to test the limits on droplet size using this much higher resolution imaging. Recent numerical computations for a Gañán-Calvo device [28] may indicate a minimum Reynolds number below which jetting cannot occur at any Weber number. If this is true of GDVN devices in general, it would clearly limit the minimum drop size.

When operated in air, the distance over which the droplets maintain a straight-line stream decreases with increasing gas pressure. This may be due to the lower inertia of smaller drops as well as increasing effect of turbulence at the higher Reynolds number [22]. When operated in vacuum at $10^5$ torr, the expanding GDVN gas quickly rarefies to the point of free molecular flow [29]. Under these conditions, the straight-line form (more exactly, the parabolic path in the gravitational field) persists indefinitely.

Slight misalignment of the liquid nozzle within the gas aperture limits our ability to further stretch the jet by increasing the gas pressure: As the gas pressure increases, the Venturi effect causes a drop in gas pressure at the side of the jet which is closer to the GDVN channel sidewall, deflecting the jet to this side to eventually attach to the sidewall (Coanda effect).

**Acknowledgements**
This research was supported by grant DBI-0555845 from the National Science Foundation, award W911NF-05-1-0152 from the Army Research Office and an NSF award from the Center for Biophotonics at UC Davis.

**References.**

[1]   U. Weierstall, R.B. Doak, J.C.H. Spence, D. Starodub, P. Kennedy, J. Warner, G.G. Hembree, P. Fromme, and H.N. Chapman, Exp. Fluids, DOI 10.1007/s00348-007-0426-8 (2007) and in press.

[2]   http://xfel.desy.de/tdr/tdr

[3]   http://www-ssrl.slac.stanford.edu/lcls/ and http://www-ssrl.slac.stanford.edu/lcls/papers/lcls_experiments_2.pdf

[4]   A. Frohn and N. Roth, **Dynamics of Droplets**, Springer, Berlin, 2000.

[5]   The primary use of electrospray sources in biological applications is for soft ionization of very large molecules pursuant to mass spectrometry. In such applications the charge-to-mass spectrum is the objective, not the actual physical structure of the species.

[6]   J.C. H. Spence and R.B. Doak, Phys. Rev. Lett., **92**, 198102 (2004).

[7]   D. Starodub, P. Rez, G. Hembree, M. Howells, D. Shapiro, H.N. Chapman, P. Fromme, P. Fromme, K. Schmidt, U. Weierstall, R.B. Doak, and J.C.H. Spence, J. Synchrotron Rad. **15**, 62-73 (2007).

[8]   D. Starodub, R.B. Doak, K. Schmidt, U. Weierstall, J.S. Wu, J.C.H. Spence, M. Howells, M. Marcus, D. Shapiro, A. Barty, and H.N. Chapman, J. Chem. Phys. **123,** 244304/1-7 (2005).

[9]   B.T. Ruotolo, K. Giles, I. Campuzano, A.M. Sandercock, R.H. Bateman, and C.V. Robinson, Science **310**, 1658-1661 (2005).

[10]  J.W.S. Rayleigh, Proc. Lond. Math. Soc. **10**, 4-13 (1878).

[11]  J.W.S. Rayleigh, Proc. Roy. Soc. Lond., **A29**, 71-97 (1879).

[12]  M. Faubel, S. Schlemmer, and J.P. Toennies, Z. Physik D - Atoms Molecules and Clusters 10, 269-277 (1988).

[13]  M. Faubel, B. Steiner, and J.P. Toennies, J. Chem. Phys. **106**, 9013-9031 (1997).

[14]  M. Faubel, B. Steiner, and J.P. Toennies, J. Electron Spectroscopy and Related Phenomena **95**, 159-169 (1998).

[15]  M. Faubel, B. Steiner, and J.P. Toennies, Molec. Phys. **90**, 327-344 (1997).

[16]  The Reynolds number is defined to be $Re=\rho U D/\mu$ where $D$ is the nozzle diameter, $\rho$ the fluid density, $\mu$ the fluid viscosity, and $U$ the flow speed.

[17]  S. Middleman, **Modeling Axisymmetric Flows,** Academic, New York, 1995.

[18]  $We = \rho U^2 D/\sigma$, where $D$ is the nozzle diameter, $\rho$ the fluid density, $U$ the flow speed, and $\sigma$ the surface tension of the fluid.

[19]  The onset of clogging at around 10 μm particle diameter may be due simply to the exponential increase in suspended atmospheric particles below this size. Actual particulate concentrations are measured and tabulated by the EPA as part of the PM2.5 and PM 10 air quality standards (see http://www.epa.gov/ttn/naaqs/).

[20]  A. M. Gañán-Calvo, Phys. Rev. Lett. **80,** 285-288 (1998).

[21]  A. M. Gañán-Calvo and A. Barrero, J. Aerosol Sci. **30,** 117-125 (1999).

[22]  S.L. Anna and H.C. Mayer, Phys. Fluids **18**, 121512/1-13 (2006).

[23]   G.L. Switzer, Rev. Sci. Instrum. **62,** 2765-2771 (1991).

[24]   M.B. Goodman and S.R. Lockery, J. Neuroscience Methods **100,** 13-15 (2000).

[25]   See, e.g., http://www.newobjective.com/products/tips_online_tips.html

[26]   A.S. Utada, L.-Y. Chu, A. Fernandez-Nieves, D.R. Link, C. Holtze, and D.A. Weitz, MRS Bull. **32**, 702-708 (2007).

[27]   J.M. Gordillo, M. Pérez-Saborid, and A.M. Gañán-Calvo, J. Fluid. Mech. **448**, 23-51 (2001).

[28]   A.M. Gañán-Calvo and M. Herrada, private communication.

[29]   H. Pauly, Atom, Molecule, and Cluster beams I, Springer, Berlin, 2000.